\documentstyle[12pt]{article}
\def\vecc{\mathaccent"017E}
\newcommand{\en}{\enspace}
\newcommand{\be}{\begin{equation}}
\newcommand{\ee}{\end{equation}}
\newcommand{\tfrac}[2]{{\textstyle{#1\over#2}}}
\author{P.B Wiegmann\\
{\it James Frank Institute and Enrico Fermi Institute }\\
{\it of the University of Chicago},
\\{\it 5640 S.Ellis Ave., Chicago IL, 60637}, \\
{\it and Landau Institute for Theoretical Physics}\\
e-mail:wiegmann@control.uchicago.edu\\}
\title{Topological Mechanism of Superconductivity}

\vspace{0.5cm}

\date{September 19, 1994}
\begin{document}\maketitle

\centerline{\it Lectures given at the XI South African Summer School in
Theoretical Physics, January 1994}
\begin{abstract}
We outline the basic ideas of the topologigical mechanisms of
superconductivity.
A gauged model of correlated electronic system where a topological fluid is
 formed as a result of a strong interaction is discussed.\end{abstract}

\newpage
%\cite{1}

\section{Citerion for Superconductivity}\label{I}
\subsection{London Equations as Hydrodynamics of an Ideal Liquid}

Our understanding of superconductivity starts from the London
equations.
They describe the linear response of superconductive matter to an
external electromagnetic field:
\begin{equation}\label{L1}
\vec{\nabla }\times \vec{j}=-{1\over{\lambda}^{2}}{\vec{B}}^{\rm ext}
\enspace ,
\end{equation}
\begin{equation}\label{L2}
\vec{\nabla}\rho +{{v_ 0}^{-2}}{\partial}_{t}\vec{j}= \kappa{\vec{E}}
^{\rm ext}\enspace .
\end{equation}

The first equation implies the Meissner effect ($\lambda$ is the
London penetration depth).  The second means that the matter is
compressible \footnote{ As a result of Coulomb interactions the liquid
is in fact incompressible.  Below we neglect the Coulomb interaction
between charge carriers and consider the electromagnetic field as
external. } ($\kappa$ is a compressibility) and implies an ideal
conductivity ($\sigma(\omega)\sim
i\omega^{-1}$).  Of course, due to gauge invariance (or the Maxwell
condition $\vec{\nabla }\times \vec E={\partial }_{t}\vec B$), the
penetration depth and compressibility are related by $\kappa = (
\lambda v_0)^{-2}$.

These two equations may be written as one.  In the Lorentz
gauge $\vec\nabla \vec{A}+v_ 0^{-2}\partial_tA_0=0$, for example,
one has
\begin{equation}
\label{L3}
\vec j=-{\lambda}^{-2}\vec{A}^{\rm ext}\en .
\end{equation}

These equations describe the hydrodynamics of the superconductor at
low temperatures.

 If we replace ${\vec{E}}^{\rm ext}$ in (\ref{L2})
by the gradient of pressure $\vec{\nabla }p$, and ${\vec{B}}^{\rm
ext}$ in (\ref{L1}) by a torque, we obtain the linearized Euler
equation for an ideal compressible liquid.  Therefore the Meissner effect is a
direct consequence of the fact that superconductive matter is a
liquid.

Let us elaborate this simple but important observation a bit further.
The London equations can also be written in terms of current-current
correlators at small $\omega$ and $k$.  Varying these equations over
${\vec{A}}^{\rm ext}$ and taking into account the continuity equation ${
\partial
}_{ t}\rho + \vec{\nabla }\vec{j}=0$, we find
\begin{equation}
\label{L4}
\langle\rho  ( k , \omega  ) \rho  (- k , \omega  )\rangle = -
\kappa {{ v_ 0}^{2} k^{ 2} \over { \omega }^ 2-{ v}_{ 0}^{2}{ k}^{ 2}}
\end{equation}
and
\begin{equation}
\label{L5}
\langle{ j}_{ \perp }( k , \omega  ){ j}_{ \perp
}( k , \omega  )\rangle =
{1\over {\lambda }^2},\qquad
\vec{ \nabla }\vec{{j}_{\perp }}=0 \en ,
\end{equation}
where $\vec j_{\perp }$ is the transverse component of the current.
Now it is obvious that our matter is a liquid.  The second equation
(which equivalently reflects the Meissner effect) implies that there
is no gapless transverse mode (transverse sound), i.e.\ that there is
no shear modulus -- this is the definition of a liquid.  A transverse
sound is a feature of a solid for which, instead of (\ref{L5}), one
has
\begin{equation}
\langle{ j}_{ \perp }(k , \omega  ){ j}_{ \perp}
( k, \omega  )\rangle = -
{1\over {\lambda }^2 } {\omega^{ 2} \over { \omega
}^ 2-{ v}_{ t}^{2}{ k}^{ 2}}\en .
\end{equation}
The density-density correlator (\ref{L4}) shows a gapless mode (longitudinal
sound) which means that the liquid is ideal and compressible, i.e.\
may flow without dissipation.

All these equations can be linked to a Hamiltonian:
\begin{equation}\label{L6}
H={m\over {2\bar\rho}}[{\vec j}^2+
v_0^2(\rho-\bar\rho)^2]+\vec A^{\rm ext}\vec j+A_0^{\rm
ext}(\rho-\bar\rho)
\end{equation}
where $\bar\rho$ is an average density.
This is  the
Hamiltonian of linear hydrodynamics.

\subsection{Landau Criterion}

To be a superfluid a system must have no dissipation.It seems
difficult to combine a soft mode in density modulation and the
absence of dissipation.
A Fermi liquid, for example, shows Landau damping,
\be
\langle{ j}_{ \perp }( k , \omega  ){ j}_{ \perp
}( k , \omega  )\rangle = -
{1 \over { \chi k^ 2+i\gamma\omega/k}}\en ,
\ee
due to gapless particle-hole
excitations ($\chi$ and $\gamma$ are constants).
  To
have no dissipation a liquid {\it must have no gapless density modes rather
than
longitudinal sound}. In particular the single particle spectrum of the
homogeneous liquid
must have a gap. In known examples it is sufficient that all other
excitations, which change a local density
 also have a gap.
This is the subject of the Landau criterion \cite{Landau}
of superfluidity.   In its grotesque
form it states that
\begin{quote}
{\it the spectrum must contain longitudinal sound and the single
particle spectrum must have a gap}.
\end{quote}

It is  believed
that this condition on the spectrum of quantum liquid is sufficient to
derive the Meissner effect.

All of this is true for a homogeneous liquid.  The true check whether
a liquid is superconductive may be made only if some non-zero
concentration of impurities does not result in resistivity.  A further
belief is that weak impurities do not lead to resistivity whenever the
Meissner effect holds in the absence of impurities.

The Landau criterion seems to be sufficient in spatial dimensions
greater than one.  In one dimension it fails for a very simple reason,
namely, in one dimension we cannot distinguish between a liquid and a
solid since there are no properties such as transversal modes and
shear modulus.  There is no Meissner effect either.
As a result  a single impurity pins down the flow in the same way as a
single impurity
 pins down a solid.

In the early days of superconductivity, Frohlich \cite{Fr} noticed
that in a one-dimensional metal an incommensurate charge density
wave (CDW) will slide through the lattice unattenuated.  Since it
carries an electric charge and since a gap has developed in the
electronic spectrum, Frohlich concluded that the ground state of his
system is superconductive.  Although, a sliding charge density wave
indeed contributes to conductivity, the Frohlich superconductivity
in one dimension was considered nothing more than a theoretical
curiosity, because of a variety of pinning mechanisms \cite {Lee}.

The failure of the charge density wave mechanism in one dimension
does not devaluate Frohlich's ideas, which may be valid in higher
dimensions \cite{Wiegmann}.  The distinctive property of Frohlich
superconductivity is {\it a topological order} -- a topological
character of the ground state.  We therefore call it the {\it
topological mechanism of superconductivity}.  One of the best known
examples of this phenomenon in two dimensions is the superconductivity
of particles with fractional statistics \cite {Laughlin} or anyon
superconductivity (see e.g.\cite{Shapere}).

There are two reasons why
the topological mechanism of superconductivity is of interest.
First of all it is a fundamentally new mechanism of superconductivity
and superfluidity.  Secondly it has been found in models of strongly
correlated electronic systems, namely in the doped Mott insulator
(some people consider it to be a mechanism of  high temperature
superconductivity).

  Below we discuss some models of topological
superconductivity and discuss why they are relevant to the physics of
 correlated electronic systems.

\subsection{Topological order}

Jumping ahead, let us discuss some features of the ground state with a
topological order.  Let the wave function of the system be
$\psi(\{x_i\})$ where $\{x_i\}$ are coordinates of particles. Its defines a
fiber bundle.
We shall refer the Chern number of this fiber bundle as a topological order.

Consider for example a two dimensional system. A topological order implies
that the ground state wave function has zeros with a
non-vanishing signature.  In this case the Chern's number is
\begin{equation}\label{T1}
Q =\sum_{a =1}^{ M}
\int_{}^{}\varepsilon {}_{ij} {\partial  \over \partial { x}_{a}^{i}}\psi
{}^*{\partial  \over \partial { x}_{a}^{j}}\psi {{d}^{ 2}{ S}_{a} \over
 8\pi }
\end{equation}
where the integral goes over the space. This equation can be also applied
for   3-dimensional systems.
In this case the integral goes over the boundary of the system.
  In one dimension, let us
define the Chern number as follows.  Consider a closed one-dimensional
space as a boundary of a domain of the complex plane and the wave
function as the value of some analytical function in the domain.  Then
(\ref{T1}) gives the Chern number.

 Perhaps the simplest
example were a topological order occurs is  particles in a magnetic field.
Then the Chern number is merely Hall conductance. However a topological
order may occur
 in correlated electronic systems even without magnetic field, but as a
result of a strong interaction.
 Thse sytems are not exotic.They were
known from the early days of quantum physics in one dimension.  They
have also recently been discovered in quantum magnetism
\cite{KL,WWZ,KW,LZ,Scripta} and intensively discussed with respect to
fractional statistics in two dimensions.
A topological order also occurs in some models of electronic systems with
strong interaction in two and three dimensions.
Some people even think that a topological order is a universal feature of
correlated quantum systems.

An illustrative example is given by the simplest solvable model of
particles in one dimension with a $1/r^2$ interaction:
\begin{equation}\label{T2}
H =-\sum_{ i=1}^N {\partial^2\over{\partial x_i^2}}
+\sum\limits_{ i,j =1}^{{ N}}
{{\lambda (\lambda-1)}\over{(x_i-x_j)^2}} \enspace .
\end{equation}
The exact wave function of this model is
\begin{equation}\label{T3}\psi  ({ x}_{ 1}....{ x}_{N}
)=\prod_{i<j}(x_i-x_j)^\lambda \enspace .
\end{equation}
This wave function tells us that each collision
contributes $\pi$ to the phase shift.  The Chern class, therefore,
is the sum of jumps of the scattering phase,
\begin{equation}\label{T4}Q = N ( N -1){\lambda\over 2}\enspace .
\end{equation}

The same property takes place in almost all
one-dimensional systems with any interaction.  The reason is that in
one-dimension any small bare interaction is effectively strong.  Below
we discuss the possibility that the topological order itself may be
developed as a result of an effectively strong interaction in any
spatial dimension.

Several peculiar properties follow immediately from the assumption of
a topological order.  For example, one-particle excitations in the
one- and three-dimensional quantum antiferromagnet have fermionic
statistics, while in two dimensions the statistics is fractional
(half fermionic).

We start from reviewing the one dimensional case were topological character
of the ground state is the most transparent.

\section {Frohlich Superconductivity -- One Dimension}\label{II}

Let us now review Frohlich's ideas (see e.g.,Ref.~\cite{Heeger,BrKr}).
We start from the Frohlich model of an {\it incommensurate}
electron-phonon system with
\begin{equation}\label{F1}H =\sum{\varepsilon }_{k} {a}_{k}^{+}{a}_{k}
+\sum\nolimits\limits_{ q}^{} {\omega }_{q} {b}_{q}^{ +}{b}_{q} + g
\sum\nolimits\limits_{ x , q}^{} {a}_{k + q }^{ +}{ a}_{q
} ({ b}_{ - q}^{ +}+{ b}_{q})
\end{equation}
where $a_{k}$ and $b_q$ are electron and phonon operators,
respectively.

In one-dimensional electron-phonon systems the Peierls
instability causes a lattice displacement
\begin{equation}\label{F2}
\langle u(x)\rangle=  {\rm Re}\;{b}_{2k_{\rm F}}{e}^{i2k_{\rm F}x}
= {\Delta  \over g} \cos{ (2{ k}_{F} x +\varphi)}
\end{equation}
and a modulation of the electronic density $\rho \sim
\cos(2{k}_{F}x+\varphi )$, where $\Delta\sim\exp(-{\rm
const}/g^2)<<E_F$ is
the value of the phonon condensate at ${2k_{\rm F}}$ and ${\varphi }$
is its phase .

Frohlich noticed that the periodic density fluctuations of electrons
are fixed only relative to the lattice and can easily travel with
some velocity such that $\rho \sim \cos(2{k}_{F}( x - vt
)+\varphi )$.  This could be compensated by changing the phase
according to $\dot\varphi =-2{ k}_{F}  v$.  Therefore the
current is $j =\rho v = v ( N{k}_{F} / \pi) $, i.e.
\begin{equation}\label{F3}
{j}_{x} =-{ N \over  2 \pi }\dot{\varphi }\en ,
\end{equation}
and from continuity
\begin{equation}\label{F4}\rho  ={ \rho }_{ 0}+
{ N \over  2 \pi }\varphi ' \en ,
\end{equation}
where $N$ is the degeneracy of an electronic state (spin, for
example).  These are Frohlich's equations.

Assuming that the charge density wave (CDW) is not pinned, i.e., $ ({1
\over { v}_{ 0}}{\partial }_{ t}^{ 2}-{ v}_{0}{\partial
}_{ x}^{ 2}) \varphi =- { E}_{x}$, we obtain the eq. (\ref{L2})
with ${\lambda }^{ -2}={ N \over 2 \pi}{v}_{ 0}$, where
$v_0={\pi\bar\rho\over mN}$ is the Fermi velocity.

Since there is a gap in the electronic spectrum, and the only gapless
mode reflects the ability of the CDW to move, Frohlich concluded that
his system is superconductive.  In fact the CDW is an ideal conductor
rather than a superconductor due to pinning mechanisms.  We already
mentioned that the Landau criterion does not work in one dimension -
there is no space to flow around an obstacle.

Let us now derive the Frohlich equations formally.  The first step is
to  pick out the fast
variables and keep only the slow variables.  In this case the slow
variables are associated with electrons in the vicinity of two Fermi
points $\pm k_{\rm F}$,
\begin{equation}\label{F5}a ( x )\sim { e}^{i{k}_{F} x} { \psi }_{L}
+{ e}^{-i k_{F}{x}}{\psi }_{R} \en ,
\end{equation}
and phonons with momentum close to
\begin{equation}\label{F6} \sum\nolimits\limits_{ q}^{}
{b}_{q}{e}^{ixq} \sim
{{\Delta } \over g} { e}^{i (2{ k}_{F} x + \varphi)}
\en .
\end{equation}
In the continuum limit we then obtain the so-called linear $\sigma
$-model
\begin{equation}\label{F7}L =
{{{\dot{|\Delta |}}^ 2} \over{g^
2{\overline \omega }^ 2}}-{{{| \Delta |}^ 2}
\over {g^ 2}}+ \overline{\psi } ( i \hat{D} -| \Delta|
e^{i\gamma _5 \varphi } ) \psi \en ,
\end{equation}
where $\hat{D} =\tilde{\gamma}_\mu({\partial }_\mu +A_\mu^{\rm ext})$
and $ \tilde{\gamma }_{0 },\tilde{\gamma }_{1 },\tilde{\gamma }_{5} $
are two-dimensional Dirac matrices, and $\overline\omega$ is a
characteristic friequency of phonons.
The modulus of the phonon field
does not fluctuate much and is determined by its mean field value
$\Delta\sim\exp(-{\rm const}/g^2)<<E_F$.  This is the Peierls
instability -- as a result of the interaction a gap $\Delta$ has
opened at the Fermi level.  Let us stress that position of the gap is
always at the Fermi level, so that the spectrum strongly depends on
the number of particles (filling factor).

Consider now a linear response to an electric field (there is no
magnetic field in one dimension), which brings us to the subject of the
axial current anomaly.  The point is that the continuum model
(\ref{F7}) possesses a local axial gauge symmetry,
$\psi(x)\rightarrow\psi(x)\exp i\gamma_5\alpha(x)$, in additional to
the ordinary gauge symmetry, $\psi(x)\rightarrow\psi(x)\exp
i\alpha'(x)$.  This would mean conservation of the axial current,
$\partial_\mu j_\mu^5={\nabla}\rho +{{v_ 0}^{-2}}{\partial }_{t}{j}=0$
as well as the ordinary current, $\partial_\mu j_\mu={ \partial
}_{ t}\rho + {\nabla }{j}=0$.  As a result of this ``symmetry'' an
external electric field may be gauged away which in it turn means that the
system does not respond to an electric field.  This of course not
true, due the axial anomaly -- in the quantum theory the axial current
is not conserved.

The axial current anomaly tells us that an external electric field
gives rise to the non-conservation of the axial current ${j}_\mu^{
5}=\overline{ \psi } {\tilde{ \gamma }}_\mu {\tilde{
\gamma }}_{ 5} \psi $,
\begin{equation}\label{F8}
{ \partial }_{\mu } { j}_{\mu }^{ 5}={ N \over
2 \pi } { E}^{\rm ext}
\end{equation}
where we have set ${v_0} = {e} = {c} = {1}$ and $ {\overline \omega}
\rightarrow {\infty}$. Using that in 1D, ${j}_{\mu }^{ 5}$ =
${\epsilon }_{\mu\nu}{j}_\nu$, we get a one-dimensional version of the
(\ref{L2}).
\begin{equation}\label{F9}
{\epsilon }_{\mu \nu } {\partial }_{ \mu }{j}_{\nu }
={ N \over  2 \pi } { E}^{\rm ext} \en .
\end{equation}

In particular this equation means that the electronic system is
compressible, even though the single particle spectrum has a gap.
Compressibility is an inherit feature of the Peierls instability in
incommensurate systems -- we already stressed that the gap always
opens at the Fermi level.  Therefore, if one adds a particle to the
system, it will not go to the upper band to occupy the lowest empty
state.  Instead it rearranges the period of the CDW, so as to create
one more level in the lower band.  The energy of the system does not
change much by adding a particle, so the system is compressible.

 This
phenomenon implements the so-called level crossing.  The fact is that
in the presence of a kink in the phase,
$\phi(\infty)-\phi(-\infty)=2\pi$, the electronic spectrum remains
unchanged, except that one level appears at the top of the lower band
with an energy $E=-\Delta$.  When we add a particle to the system, it
will therefore create a kink in the spatial configuration of $\phi$
and an extra level to be absorbed in the lower band.  The anomaly
equation (\ref{F9}) reflects this phenomenon.  We derive this
well-known result further on.

Strictly speaking, we cannot distinguish between solid and liquid in
one dimension -- there is no transversal current.  Nevertheless, there
is a global version of the Meissner effect -- a current in a closed
wire must satisfy
\begin{equation}
\int { j}dx =-{1 \over {\lambda }^{ 2}}{ \Phi}^{\rm ext}
\end{equation}
where $\Phi^{\rm ext}$ is the magnetic flux within the wire.

Combining (\ref{F8},\ref{F9}) with the Frohlich equations
(\ref{F3},\ref{F4}) and using the relation ${j}_{\mu }= -{\partial L /
\partial {A}_{\mu }^{\rm ext}}$ we obtain a bosonized version of the
incommensurate CDW:
\begin{equation}\label{F10}
L_{\varphi }= {N \over {4 \pi}} \Bigl( \tfrac{1}{2}({\partial
_ m }{\varphi })^2-{2 \over  \pi }  E^{\rm ext}\varphi \Bigr)
\en .
\end{equation}

Let us now turn to the commensurate CDW and consider an electronic
system close to a half-filled band.  At exact half filling, the CDW is
two-fold commensurate and so the vacuum is two-fold degenerate.  A
canonical tight-binding model is
\begin{equation}
{H} =\sum\nolimits\limits_ n {\Delta }_{ n , n
+1}({ a}_{n}^{ +}{ a}_{n +1} +h.c)+{H}_{\Delta }
\end{equation}
where ${\Delta }_{ n , n + 1}$ is a
fluctuating
hopping amplitude  and ${H}_{\Delta}$ is a
phonon energy.   In the continuum limit the half-filled case is described by
the
same ${\sigma}$- model (17) but with a real ${\Delta}$:
\begin{equation}\label{F13}L =
{\overline{ \psi }} ( i \hat{ D} - i { \gamma }_{ 5}
\Delta  \left({ x}\right) ){ \psi } -{{ \Delta }^{2} \over
{ g}^{ 2}}
\end{equation}
In this case there is no soft translational mode since the CDW is
commensurate and is pinnned by the lattice.  The excitations are kinks
of $\Delta\left({ x}\right)\ $ which connect two-fold degenerate
mean field vacua: $ \Delta \rightarrow \pm {\Delta }_{ 0} $ when $ x
\rightarrow \pm \infty $.  In the presence of the kink the electronic
spectrum remains approximately unchanged, except for appearance of
the
so-called zero mode, a state with a zero energy, located exactly in
the middle of the gap.  This mode is $1/2$ degenerate ( the index
theorem), i.e.\ it can accommodate one-half of a particle.  Fractional
degeneracy
means that if the particle has an isospin, only one particle with
spin, say up, can occupy the zero mode.  In coordinate space the
wave function is located in the core of the kink.  The axial anomaly
in this case tells us that the density of extra states is equal to the
density of zeros of $\Delta(x)$ (see, e.g.,\cite {NS})
\begin{equation}\label{F14}
\rho(x)=\tfrac{1}{2}\delta(\Delta(x))
{{\partial\Delta}\over{\partial x}} \en .
\end{equation}
Suppose now we dope the system by adding additional particles $\delta$.  The
system will lower its energy by creating the number of solitons (zeros
in $\Delta (x)$) which is necessary to absorb all dopants.

Due to the interaction between solitons they form a periodic lattice.   As a
result of that, zero modes develop a narrow band in the middle of the gap with
a width of the order
$\Lambda  \sim { \Delta }_{ 0} \exp (- {\rm const} \delta \rho/ \bar\rho) $
where $\delta\rho$ is the doping density and $\bar\rho$ the  density of the
undoped system.
This band absorbs all the dopants and is always completely full.

As in Frohlich's case, these solitons have a translation mode due to
their topological origin: a soliton lattice can slide along the atomic
lattice without dissipation.  Let $\overline x_i$ be the zeros of
$\Delta (x)$.  Then the
density of extra particles (dopants) is
$\delta \rho ( x )={1\over 2}\sum_{ i} \delta (
x -{\overline x}_{i})$.  Displacement of the positions of zeros around
their mean field values, ${x}_{i} =\overline{{ x}_{i}} + \varphi ({
x}_{i} , t )/2 \pi $, give rise to fluctuations of the density
$\delta\rho (x)= \rho ( x )-\overline{ \rho }$.  According to
(\ref{F14}) they obey the same Frohlich equations (\ref{F3},\ref{F4}),
 \begin{equation}\label{F15}
{j}_{\mu } ={\overline\rho\over {2\pi}}
{ \epsilon }_{\mu  \nu } {\partial }_{\nu } \varphi \en .
\end{equation}
The anomaly equations (\ref{F9}), with an extra $\overline \rho$,
apply again\footnote{
In the lower density limit the factor $\overline \rho$
is hidden in the velocity $v_0$.  Here and later on  we neglect its
space dependence.}

\begin{equation}\label{F16}
{ \epsilon }_{\mu  \nu } {\partial }_{\nu
}j_\mu={\overline{ \rho }\over 2}E^{\rm ext} \en .
\end{equation}
Each twist of $\varphi$ adds one additional state in the middle of the
gap.  Therefore, adding $n_e$ additional particles gives rise to the
topological charge of $n_e = Q$ where $Q = \int {d\varphi\over 2\pi}$.
All of this is true in incommensurate cases when the system, after
doping, has infinitely degenerate classical vacua.  If the doping is a
rational number, say, $p/q$, then the number of degenerate vacua is
finite, namely $q$.  The CDW is pinned, generally by an exponentially
small potential ${\Delta }_{ 0}^{2}({ \Delta }_{ 0}/{ \varepsilon
}_{f} ){}^{ q -2} \cos q\phi$.

\medskip

\noindent Let us list some important messages that follow from this
picture:
\begin{enumerate}
\item[(i)] An energy gap in the electronic spectrum eliminates elastic
scattering.

\item[(ii)] The system has infinitely many degenerate vacua.   They
are characterized
by the number of solitons, i.e., by a topological charge $|Q\rangle$.

\item[(iii)] A topological configuration $\varphi(x,t)$ such as
$\varphi
(x,t = 0) = 0$ and $\int \partial_{x}\varphi {dx} =2 \pi $ at $t
\rightarrow \infty $, which transforms one vacuum $ |Q\rangle$ into
another $ |Q + 1\rangle$, is a low energy excitation with a dispersion
$\omega(k) = v_o k$.  Then ``the whole system, electrons and solitons,
can move through the system without being disturbed'' \cite{Fr}.

\item[(iv)] This, however, is not sufficient in dimensions greater
than
one. To achieve superconductivity in higher dimensions, it must have
all these properties {\em and in addition be a liquid}.
\end{enumerate}

\section{Fermionic Number and Solitons}\label{III}
\subsection{Level Crossing}

Let us consider fermions in a static (vector or scalar) potential.
When the potential changes adiabatically, the fermionic energy levels
also shift.  If the chemical potential $\mu$ lies in the gap, then for
an adiabatic change of the potential the levels generally cannot cross
the energy level $E=\mu$ and therefore
all occupied levels remain below $\mu$.  However, there are certain
potentials which create some unoccupied levels below
$\mu$ or force some occupied levels to cross the level of the chemical
potential.  This phenomenon is known as level crossing.  Unoccupied
levels appear below (or occupied levels appear above) the chemical
potential and separated from the continuous part of the spectrum are
what is called zero modes.

There is an important theorem (the index theorem) which states that
potentials which give rise to crossing levels, must have a topological
charge.  We refer to them below as solitons.  In Sect.\ \ref{II} we
discussed solitons (kinks) in one-dimensional models.  Below we
review them and discuss solitons in some models in 2 and 3 dimensions
which will illustrate the essential physics of the topological
mechanism.

\subsection{Models and Solitons}
\subsubsection{ One Dimension -- Kinks.}

The most general Hamiltonian of the Peierls model is
\begin{equation}\label{s1}
H=\alpha_xi\partial_x+\beta
\pi_1+i\gamma_5\pi_2
\end{equation}
where the Dirac matrices $\alpha_x,\beta,\gamma_5=\alpha\beta$ may
chosen
as the Pauli matrices:
$\alpha_x=\sigma_3,\beta=\sigma_1,\gamma_5=-i\sigma_2$.
We also assume that the modulus of the vector $(\pi_1,\pi_2)$ takes a
fixed value at infinity.

The soliton  here is a field $\vec\pi(x)$ which forms a homotopy class
$\pi_1(S^1)$.  Its topological charge is
\begin{equation}\label{s2}
Q=-\int{{\vec\pi\times\partial\vec\pi}\over {\vec\pi\cdot\vec\pi}}
{dx\over 2\pi}
\en.
\end{equation}
 With a given value of the $|\vec\pi|$ at infinity $Q$ is an integer number.

\subsubsection{Two Dimensions -- Vortices.}

In dimensions greater than one it is necessary to include {\it a vector
gauge field} in addition to a scalar field.

\noindent  We consider two models:

A.  QED$_3$
\begin{equation}\label{s3}
H=\sum_{\sigma=1,2}\psi^{\dag}_\sigma[\alpha_\mu(i\partial_\mu+a_\mu)+\beta
m]\psi_\sigma  \en .
\end{equation}

A soliton for this model is a vertex or magnetic flux with the charge
\begin{equation}\label{s4}
Q=-\int {{\vec\nabla \times\vec a}\over 2\pi}dx \en .
\end{equation}

B.   Another model is non-abelian (SU(2)) QCD$_3$ coupled to a meson
field
\begin{equation}\label{s5}
H=\alpha_\mu(i\partial_\mu+\vec
A_\mu\vec\tau)+\beta\vec\tau\vec\pi \en .
\end{equation}

Solitons in this model are a combination of the hedgehog of the meson
field (an element of the class $\pi_2(S^2)$) and a vortex of the gauge
field.  Its topolological charge is
\begin{equation}\label{s6}
Q=\int[-{\vec\pi \vec F \over
|\vec\pi|}+{1\over4}{\vec\pi\cdot D\vec \pi\times D\vec \pi\over
|\vec\pi|^3}]{d^2x\over 4\pi}
\end{equation}
where $\vec F=\partial_1\vec A_2-\partial_2\vec A_1+\vec
A_1\times\vec A_2$ and $D\vec\pi=
\partial\vec\pi+2\vec A\times\vec\pi$ is the covariant derivative.
A soliton with
a topological charge of 1  has the asymptotic form
\begin{equation}
\label{s7}\pi_3=0\qquad( \pi_1,\pi_2 )\rightarrow {\vec{r} \over
r}\qquad {A}_{i}^{3}\rightarrow {1 \over 2}{\varepsilon }_{ij}{{r}_{j}
\over {r}^{2}} \enspace .
\end{equation}
at infinity.
A general form of a vortex with an arbitrary topological charge is given by
the homotopy class $\pi_2 (S^2)$.

Let $ \hat{\pi}=\vec{\pi }\vec{\tau }/|\vec{\pi }|$ be an element of
this class.  We shall set
\begin{equation}\label{s8}
{ A}_{i}={1 \over 2}{\hat{\pi }}{\partial }_{i}{\hat{\pi}}
={\vec{A}}_{i}{\vec{\tau }} \en ,
\end{equation}
a vortex with a charge
\begin{equation}\label{s9}
Q={1 \over {8\pi}}\int {\rm tr}(\hat{\pi }d\hat{\pi}d\hat{\pi}) \en .
\end{equation}
 A unit of the topological charge  is carried by a zero
of $\vec \pi$.  The density of the topological charge is given by
the ``magnetic
field'' projected onto $\vec\pi$
\begin{equation}\label{s10}
B^{\rm vortex}\equiv
\hat{\vec{ \pi }}{\vec{F}}^{\rm vortex}=\tfrac{1}{2} {\varepsilon}^{abc}
{\hat{\pi }}^{a}{\partial}_{1}{\hat{\pi }}^{b}{\partial }_{2}{\hat{\pi }}^{c}
\end{equation}
so that $Q={1 \over {2\pi}} \int B^{\rm vortex}d{\vec x}$ is a total flux.

\subsubsection{Three Dimensions -- Monopoles.}

As a model in three spatial dimension we consider the SU(2) gauge
theory with chiral bosons:
\begin{equation}\label{s11}
H=  \alpha_\mu(i\partial_\mu+\vec
A_\mu\vec\tau) + i{\gamma _ 5}{\vec\pi}{\vec\tau} + m \en .
\end{equation}
 The spectrum of this model is symmetric since the matrix $\gamma_5$
anticommutes with the Hamiltonian.

 Solitons for this model are  the Polyakov - 't Hooft
monopoles.   At  $m = 0$, and at $ r \rightarrow \infty $ the monopole
configuration is
\begin{equation}\label{12}
\vec{ \pi }(r)\rightarrow {\vec{r} \over r}\qquad
{A}_{i}^{a}\rightarrow \tfrac{1}{2}{\varepsilon }_{iaj}{{r}_{j} \over
{r}^{2}} \en .
\end{equation}

Similar to the two-dimensional case the soliton configuration is given
by (\ref{s8}) and is an element of ($\pi_2 (S^2)$).  Then the monopole
charge is carried by zeros of $\vec \pi$ \cite{F}.  It has the form of
(\ref{s9}) where the integral now goes over a boundary surrounded
 zeros of $|\vec\pi|$.  The density of the
monopole's charge is
\begin{equation}\label{s13}
q= {1 \over {8\pi}}\varepsilon_{ijk}{\rm tr}(\hat{\pi }
\partial_i\hat{\pi}\partial_j\hat{\pi })\nu_k
\end{equation}
where
\begin{equation}\label{s14}
\nu_k=\partial_k|\vec \pi|\delta(|\vec
\pi|)
\end{equation}
is the density of zeros of $\vec \pi$.

The ``magnetic field'' projected onto $\vec\pi$ is a gauge invariant,
\begin{equation}\label{s15}
B_{i}^{\rm mon}\equiv \tfrac{1}{2}{\varepsilon }_{ijk} \hat{\vec{
\pi }}{\vec{F}}_{jk}^{\rm mon}= {\varepsilon }^{abc}
{\varepsilon }_{ijk}{\hat{\pi
}}^{a}{\partial }_{j}{\hat{\pi }}^{b}{\partial }_{k}{\hat{\pi }}^{c}
\en.
\end{equation}
So
$q={1 \over {4\pi}}\nu_i B_{i}^{\rm mon}$,
which is equivalent to the Dirac equation for the monopole charge
$q={1 \over {4\pi}}\partial_iB_{i}^{\rm mon}$.

Rotation of the vector $\hat{\vec{\pi}}$ implies a gauge transformation
\cite{F}.  It is convenient to describe a multimonopole system
 in the {\it unitary gauge}.
If $g$ is a transformation which rotates $\hat{\vec{\pi}}$ to the
third axis, $ \hat\pi = g^{-1} \tau_3 g $,
then the gauge transformation
\be
\hat\pi \rightarrow g \hat\pi g ^{-1} = \tau_3
\ee
\be
A_i \rightarrow g A_i g^{-1} - \partial_i g g^{-1} =  A_i^D \tau_3
\ee
converts the non-abelian monopole to the Dirac monopole:
\be
A_i^D = -{\rm tr} (\tau_3 \partial_i g g^{-1})\en .
\ee

This gauge is singular.  It transfers a topological charge from
the $\vec\pi$-field to the gauge field.  A general form of the
topological charge can be written in a compact form if one includes
the mass $m$ into the 4-vector $\pi_a=(m,\vec\pi)$:
\begin{equation}\label{s16}
Q = {1 \over {2\pi ^2}}\int\varepsilon
_{ijk} {\varepsilon _{abcd}}[{\pi_a D_i{ \pi}_b D_{j}{ \pi}_c D_{k}{
\pi}_d \over |\vec\pi|^4 } -{3\pi_a D_{i}{ \pi}_b F_{jk}^{cd}\over
{4{|\vec \pi|}^2}}]d^3x \en .
\end{equation}

\subsection{Anomalous Calculus}

In this section we calculate the so-called ``fermionic number'' or the
Atiah-Patodi-Singer invariant $\eta$ for the models discussed in the
previous section.  It is simply the number of new states appearing
above (or below) a certain energy when one switches on the potential.
For Dirac particles it is the number of states appearing inside the
gap.  These states are refered to as zero modes or midgap states.
Since without a potential the spectrum of the Dirac operator is
symmetric, this number also measures the difference between the number
of states with positive and negative energy $\eta=N_+-N_-$ (spectral
asymmetry).  If $N=N_++N_-$ is the total number of levels, then the
number of negative levels would be $N_-={N\over 2}-{1\over 2}\eta$.
The ``regularized'' number of negative levels i.e.\ the change of the
negative levels due to the potential, $\Delta N=N_ -{N\over 2}=-{1\over
2}\eta$, is the fermionic number.

The index theorem (see e.g.\ \cite{NS} states that the fermionic
number
of the Dirac operator, or the number of extra states induced by a
soliton, equals the topological charge of the soliton, $\Delta N=Q$.
Moreover the density of the extra states is equal to the topological
density $q(x)$:
\be
\rho(x)\equiv(\psi^{\dag} (x)\psi(x)-{1\over
2})=q(x) \en.
\ee
We illustrate this general fact for the models listed in
this section.

The fermionic number for the Dirac operator may be defined
as
\begin{equation}
\label{a1}\Delta N=-{1\over 2}{\rm Tr}[{\rm sgn}\; H] \en .
\end{equation}
This divergent quantity has to be regularized.  One of the standard
ways of regularization is to replace it by
\begin{equation}\label{a2}
\Delta N=-{1\over 2\pi}\int{\rm Tr}[{H\over
{H^2+z^2}}]
\end{equation}
The integrand now converges.

\subsubsection{One Dimension -- Kinks.}
Let us start from the Peierls model (\ref{s1}),
\begin{equation}\label{a100}
H=\alpha_xi\partial_x+\beta
\pi_1+i\gamma_5\pi_2\qquad\alpha_x=\sigma_3,\; \beta=\sigma_1,\;
\gamma_5=-i\sigma_2 \en .
\end{equation}
Then the square of the Hamiltonian is
\begin{equation}\label{a3}
{H}^2=(i\partial_x)^2+\vec{\pi}^2+\varepsilon_{ab}\sigma_a\partial_x{\pi}_b,
\qquad
a,b=1,2 \en .
\end{equation}
Let us now expand the integrand of (\ref{a2}) in
$\partial\vec{\pi}$.  In the first non-vanishing
order we have
\begin{equation}\label{a4}
\Delta N =-\frac{1}{2\pi}
\int_{-\infty}^{+\infty}\;
{\rm Tr}\left( \frac{\sigma^3i\partial+\vec{\sigma}\vec{\pi}}
{(i\partial)^2+\vec{\pi}^2+z^2}i\sigma^3\vec{\sigma}
\partial\vec{\pi}
\frac{1}{(i\partial)^2+\vec{\pi}^2+z^2}
\right)
\end{equation}
This is the only order which contributes to the fermion number.
The trace in spinor space gives
\begin{equation}
\Delta N =-\frac{1}{\pi}
\int_{-\infty}^{+\infty}dz\;
{\rm Tr}\left(\frac{\pi^1\partial\pi^2-\pi^2\partial\pi^1}
{[(i\partial)^2+\vec{\pi}^2+z^2]^2}\right) \en .
\end{equation}
The trace in momentum space is conveniently calculated in the
plane wave basis,
\begin{equation}
\Delta N =-\frac{1}{\pi}
\int_{-\infty}^{+\infty}dz\; \int dx\; \int \frac{dp}{2\pi}
\epsilon_{\mu\nu}\pi^{\mu}\partial\pi^{\nu}
\frac{1}{[p^2+\vec{\pi}^2+z^2]^2} \en .
\end{equation}
Finally the integrals over $p$ and $z$ give
\begin{equation}\label{a10}
\Delta N =-\frac{1}{2\pi}
\int dx\;
\epsilon_{\mu\nu}\frac{\pi^{\mu}\partial\pi^{\nu}}{\vec{\pi}^2} \en ,
\end{equation}
the topological charge (\ref{s2}) of  kinks.
The number of extra states induced by a soliton therefore equals
the topological charge of the soliton.

The model with $\pi_1=0$ deserves special interest. This is
commensurate Peierls model ($\pi_2=\Delta$). It is defined by
\begin{equation}
\label{a11}H=\alpha_xi\partial_x+i\gamma_5\Delta \en .
\end{equation}
The spectrum of the Hamiltonian is symmetric.  This means that  if there
is a state $\psi_E$ with an energy $E$ then there is always a state
$\beta\psi$ with the
energy $-E$, except for $E\ne 0$. It follows from the fact that the
Hamiltonian anticommute with the matrix $\beta$.

Now the soliton is a kink  with
$\Delta(-\infty)=-\Delta(+\infty)=\Delta_0$.
Setting $\pi_1=0$ in (\ref{a10}), we obtain (\ref{s14}) in the form
\begin{equation}\label{a12}
Q=-{1\over2} \int\delta(\Delta)\partial\Delta
dx \enspace .
\end{equation}
In this case an extra state appears in the middle of the gap (zero mode).
 The kink does not respect periodic boundary conditions. As a result of
this a zero mode state
 may accommodate only 1/2 of the particle. In a system with a periodic
boundary conditions kinks may
appear only together with antikinks, so the total number of states remains
integer.

\subsubsection{Two Dimensions -- Vortices.}
\paragraph{A. QED$_3$.}

We now consider the model (\ref{s3})
\begin{equation}
H=\sum_{\sigma=1,2}\psi^{\dag}_\sigma[\alpha_\mu(i\partial_\mu+a_\mu)+\beta
m]\psi_\sigma  \en .
\end{equation}.  The square of its
Hamiltonian is \begin{equation}
{\cal H}^2=(i\partial_{\mu}+a_{\mu})^2+m^2-\beta(\partial_1a_2-\partial_2a_1)
\end{equation}
Let us  expand the integrand in  (\ref{a2}) in terms of
$\beta(\partial_1a_2-\partial_2a_1)$.  The first and only
non-vanishing order gives:
\begin{eqnarray}
&\Delta N =-\frac{1}{\pi}
\int_{-\infty}^{+\infty}dz\;
{\rm Tr}&\left(\frac{\alpha_{\mu}(i\partial_{\mu}+a_{\mu})+\beta m}
{(i\partial_{\mu}+a_{\mu})^2+m^2+z^2}\beta(\partial_1a_2-\partial_2a_1)
\times\right. \nonumber\\
&&\left.\times\frac{1}{(i\partial_{\mu}+a_{\mu})^2+m^2+z^2}\right)
\en .
\end{eqnarray}
The trace leaves the integral over $p$ and $z$:
\begin{equation}
\Delta N =-\frac{2}{\pi}
\int_{-\infty}^{+\infty}dz\; \int d^2x\; \int \frac{d^2p}{(2\pi)^2}
\frac{m(\partial_1a_2-\partial_2a_1)}
{[(p_{\mu}+a_{\mu})^2+m^2+z^2]^2} \en .
\end{equation}
As a result we obtain (compare (\ref{s4})) the relation between the fermion
number and the topological charge of the vortices
\begin{equation}\label{a15}
\Delta N =-\frac{1}{2\pi} \; {\rm sgn}\; m \;
\int d^2x\;
(\partial_1a_2-\partial_2a_1) \en .
\end{equation}

\paragraph{B. QCD$_3$ coupled with a meson field.}

This model is defined by
\begin{equation}\label{a20}
{\cal H}=\alpha_{\mu}(i\partial_{\mu}+\vec{A}_{\mu}\vec{\tau})+\beta
\vec{\pi}\vec{\tau}
\end{equation}
and the square of the Hamiltonian is
\begin{equation}
{\cal H}^2=(i\partial_{\mu}+\vec{A}_{\mu}\vec{\tau})^2
-\beta \vec{F}\vec{\tau}+
(\vec{\pi})^2+\alpha_{\mu}\beta \; i\partial_{\mu}\vec{\pi}\; \vec{\tau}+
\alpha_{\mu}\beta i[\vec{A}_{\mu}\times \vec{\pi}]2\vec{\tau} \en .
\end{equation}

To simplify calculations we first drop the gauge field.  Then,
expanding the integrand in  (\ref{a2})
in terms of
$\alpha_{\mu}\beta \; i\partial_{\mu}\vec{\pi}\;\vec{\tau}$ up to the second
term, we have
\begin{eqnarray}\label{a30}
&&\Delta N
 =  -\frac{1}{2\pi}
\int_{-\infty}^{+\infty}dz\;
{\rm Tr}\left[\frac{\alpha_{\mu}i\partial_{\mu}+\beta
\vec{\pi}\vec{\tau}}
{(i\partial_{\mu})^2+(\vec{\pi})^2+z^2}\times \right. \nonumber \\
&& \left(\left.\!\!
1\!-\!(\alpha_{\mu}\beta i\partial_{\mu}\vec{\pi}\;\vec{\tau})
\frac{1}{(i\partial_{\mu})^2+(\vec{\pi})^2+z^2}\!
+\!\left(\!(\alpha_{\mu}\beta i\partial_{\mu}\vec{\pi}\,\vec{\tau})
\frac{1}{(i\partial_{\mu})^2+(\vec{\pi})^2+z^2}\right)^2
\right)\!\!\right] \nonumber \\
&&
\end{eqnarray}
Changing $i\partial_{\mu}$ to $p_{\mu}$ and taking the trace  in the spinor
and isospinor
spaces we get
 \begin{equation}
\Delta N =-\frac{1}{2\pi}
\int_{-\infty}^{+\infty}\!\!dz \int\! d^2x\; \int\!
\frac{d^2p}{(2\pi)^2}
\frac{\alpha_{\mu}p_{\mu}+\beta \vec{\pi}\vec{\tau}}
{[(p_{\mu})^2+(\vec{\pi})^2+z^2]^3}i\beta
\left[
\partial_1 \vec{\pi}\vec{\tau},\;\partial_2 \vec{\pi}\vec{\tau}
\right] \en .
\end{equation}

Finally,  integrating over $p$ and $z$ we obtain the fermion number
\begin{equation}
\Delta N =\int\!\rho (x) d^2x
\end{equation}
with the density
\begin{equation}\label{a21}
\rho =\frac{1}{8\pi}
\epsilon^{ij}\epsilon^{abc}
\frac{\pi^a \partial_{i} \pi^b \partial_{j} \pi^c}{|\vec{\pi}|^3}
\en .
\end{equation}
Once again we obtain that the density of the extra states is equal
to the topological charge of the soliton.

To include the gauge field one must first replace derivatives in
(\ref{a21}) by covariant derivatives.  Then we notice that for
$\pi_1=\pi_2=0$ and $A_\mu^1=A_\mu^2=0$ the fermion number must
reproduce the QED$_3$ result(\ref{a15}). Taken together these
considerations give
\begin{equation}\label{a31}
\rho
=-\frac{1}{2\pi}\frac{\vec{\pi}\vec{F}}{|\vec{\pi}|}
+\frac{1}{8\pi}
\epsilon^{ij}\epsilon^{abc}
\frac{\pi^a
(\partial_{i} \pi^b+2[\vec{A}_{i}\times\vec{\pi}]^b)
(\partial_{j} \pi^c+2[\vec{A}_{j}\times\vec{\pi}]^c)}
{|\vec{\pi}|^3}
\end{equation}
where $\vec{F}=\partial_1\vec{A}_2-\partial_2\vec{A}_1
+[\vec{A}_1\times \vec{A}_2]$.

A special case of interest is $\pi_3=0,A_\mu^1=A_\mu^2=0$.  In this
case the spectrum is even as the matrix $\beta\tau_3$ anticommutes
with the Hamiltonian.  Equation (\ref{a31}) gives
\begin{equation}
\rho=-(\varepsilon_{ij}\partial_i
\pi_1\partial_j \pi_2+{1\over 4\pi}|\vec\pi|^2
\vec\nabla\times\vec A^3)\delta (\vec\pi) \en .
\end{equation}
In this case all extra states are located exactly in the middle
of the gap (zero modes).  Each zero mode may accommodate 1/2 of a
particle.

\subsubsection{Three Dimensions -- Monopoles.}

The fermion number for the model (\ref{s11}) is again equal to the
topological charge of the monopole (\ref{s16}) \cite{GW}.  The
simplest
way to calculate it is to drop the gauge field and proceed similarly
as in the previous section.  Now we must keep the third order term in
the expansion (\ref{a30}).  As a result we obtain
\be
\Delta N= {1 \over  {2\pi
^2}}\int\! d^3x\varepsilon _{ijk} {\varepsilon _{abcd}}{\pi_a d_i
{\pi}_b d_{j}{ \pi}_c d_{k}{ \pi}_d \over |\vec\pi|^4 }\en .
\ee
Next we want to make this expression gauge invariant.  After a
replacement of derivatives by covariant derivatives, it still remains
gauge non-invariant.  To rectify this one must add the second
term in (\ref{s16}).

Similar to the previous examples a very special case arises when
$m\equiv\pi_0=0$.  In this case the matrix $\beta\gamma_5$
anticommutes with the Hamiltonian, and the spectrum is symmetrical.
Therefore the extra states appear in the middle of the gap at $E=0$
(zero modes).  As usual each zero mode is 1/2 degenerate.  Setting
$\pi_0=0$ in (\ref{s16}) we obtain
\begin{equation}\label{a40}
\rho(x)= {1 \over {8\pi}}\varepsilon_{ijk}{\rm tr}(\hat{\pi }
[\partial_i\hat{\pi}\partial_j\hat{\pi }+{1\over 4\pi}F_{ij}])\nu_k
\end{equation}
where $\hat\pi=\vec\pi /|\vec\pi|$ and
$\nu_k=\partial_k|\vec \pi|\delta(|\vec
\pi|)$ is the density of zeros of $\vec \pi$.

\section{Topological Fluid as a Superfluid}\label{IV}
\subsection{The Basic Model -- Holons and Spinons -- Zero Current Theory}

The main inspiration for considering topological superconductivity
comes from strongly correlated electronic systems.  They suggest a
general model to study this phenomenon.  Running ahead we discuss this
model now and later derive it from a canonical model of the doped
Mott insulator.

Consider two sorts of particles ``holons'' $h$ and ``spinons'' $\psi$ (the
names come from the Mott insulator literature) which obey the so called
"zero current" theory.
This means that densities and currents obey the local constraint
:
\begin{enumerate}
\item[(i)] their densities are always complementary,
$h^{\dag}(x)h(x)=\psi(x)\psi^{\dag}(x)$,
namely the number of holes in spinons is the number of holons, where
the total number of holons is given by
$\int h^{\dag}(x)h(x)dx=\delta$ (doping)
and
\item[(ii)] their currents are always opposite (the so-called
``zero current theory'')  $\vec j_h(x)+\vec j_s(x)=0$.
\end{enumerate}
The holons carry an electric charge and spinons are neutral, so
that the holon current is
in fact the electromagnetic current.  The Hamiltonian of the basic
model is
\begin{equation}\label{m1}
{\cal H}={1\over m}h^{\dag}(x) H^2h(x)+ \psi^{}(x)
H \psi(x)+\vec A^{\rm ext}\vec j_h \en ,
\end{equation}
where $H$ may be any Hamiltonian which exhibits a level crossing.
 For example, one may consider any
one of the models considered in the previous section.  The local constraints
are
automatically enforced if $H$ contains an abelian gauge field, as in
(\ref{s3}) of QED$_3$ , other wise a gauge field ( Lagrangian
multiplier) must be added to (\ref{m1}): $\vec A(\vec j_h+\vec j_s)
+A_0(h^{\dag}(x)h(x)-\psi(x)\psi^{\dag}(x))$.  Later we derive the
Hamiltonian $H$ for the Mott insulator.  This Hamiltonian is in fact
different from any of the standart models listed in the Sect.\ \ref{III},
but has essential
 common features with models of the sec.3.

\subsection{Topological Instability}

Let us consider the basic model of topological fluids (\ref{m1}).  We
argue that the perturbative vacuum of the model is unstable with
respect to the creation of topological charge and solitons, namely we
show that the vacuum has a topological charge such that the
fermionic number equals the number of dopants (holons),
\begin{equation}\label{t1}
-{1\over 2}\eta=\Delta N = \delta\en .
\end{equation}
This the topological charge of the ground state.

To see this, let us apply the adiabatic strategy: first consider all
potentials $\vec A$ and $\pi$ static and find the best spatial
configuration to minimize the energy for a given doping, and then
consider fluctuations around the minimal static configuration.  First
assume that there are no solitons and start adding dopants.  Consider
first the spinon part of the model.  Dopants start to fill the upper
band of the Dirac spectrum of spinons.  This costs the energy of the
gap, plus the Fermi energy of spinons, plus some small radiative
corrections of order of ${ \delta} $.  If, however, there are enough
solitons to absorb by their zero modes all dopants, we pay less than
the gap and gain all that energy.  In addition we gain the Fermi
energy, since the zero modes form a narrow band inside the gap.  The
same arguments hold for the holon part of the Hamiltonian as well.
The solitons form a narrow band for the holons at zero energy, so they
also favour the topological vacuum.  As a result the number of
solitons in the vacuum must provide that number of zero modes which
will absorb all dopants, $ h^{\dag}h=\delta$.

To formalize these arguments, one can show that the perturbative
vacuum is indeed unstable.  The topological condensate appears in the
one-loop correction.  In leading order of the chemical potential, the
free energy is
\begin{equation}     E  - 1/2| \mu \eta |-\mu \delta
\end{equation}
where $E$ is a perturbative energy.  Equations (\ref{t1}) minimize
this free energy.

We conclude that a small doping requires a configuration of $\pi$ and
$\vec A$ with the relevant topological charge determined by
(\ref{t1}).  In turn the wave function of the state with a
topological charge establishes the Chern class (7) equal to the
charge. We refer to this phenomenon as topological instability.

\subsection{Topological Field Theory coupled with the Matter Field and
Superfluidity}

A simple version of the basic theory occurs when the gap in the Dirac spectrum
is the largest parameter in the system.  Integrating over spinons we
can then
neglect all terms except the so called the Wess-Zumino-Witten term,
denoted WZW hereafter. It is
the only term which does not vanish when the gap approaches infinity,
and is given by
\begin{equation}
{\rm WZW}=\lim_{gap\rightarrow\infty}\ln {\rm Det}(i\partial_t+H)
\end{equation}
The WZW term captures in the Lagrangian language the phenomenon of
anomalies, namely, anomalous
currents  may be obtained as the Lagrangian equation $j_\mu=-\delta
{\rm WZW}/\delta A_\mu$.  For the 1D model (\ref{s1}), the topological
current (see (\ref{s2}))
\be
j_\mu=-{1\over{2\pi}}\varepsilon_{\mu\nu}
\varepsilon_{ab}{{\pi^a\partial_\nu\pi^b}\over {|\vec\pi|^2}}\en
\ee
Therefore,
\be
{\rm WZW}=-\int{dx\over{2\pi}}\varepsilon_{\mu\nu}
\varepsilon_{ab}A_\mu{{\pi^a\partial_\nu\pi^b}\over {|\vec\pi|^2}}\en
\ee,
The WZW term  for QED$_3 $ it is known as the Chern-Simons term:
\be
{\rm WZW}=-\int {dx\over{4\pi}}{\rm sgn}\; m \varepsilon_{\mu\nu\lambda}
A_\mu d_\nu A_\lambda \en .
\ee
For QCD$_3$ and for the 3D model ${\rm WZW}=-\int A_\mu q_\mu dx$,
where $q_\mu$ is the topological current given by the integrand of
(32) and (42) \cite{Witten}.

The WZW term itself gives rise to and is descriptive of topological
field theories.  A topological field theory describes only the zero
mode sector of a field theory and the zero mode sector is invariant
under all diffeomorphisms of the volume -- positions of the solitons
are not fixed unless they interact with matter.  This means that any
external perturbation, such as inhomogeneous pressure or twist, does
not change the energy of the system.  As a result the zero mode sector
is topological in the sense that it has no physical states in the
bulk.  Physical states appear only as a result of a boundary and they
are confined to a boundary (the edge states).

Our basic model is a topological field theory coupled with
matter (holons),
\begin{equation} L=\int dx~ h^{\dag}(i\partial_t+A_0+H^2)h
+{\rm WZW} \en .
\end{equation}

In this case the zero modes are occupied by particles (holons) and
possess some dynamics.  Nevertheless, this dynamics is limited as a
result of the overall gauge symmetry of the unperturbed topological
field theory.  Matter of course destroys all diffeomorphisms which
change the density of the matter, so that only diffeomorphisms which
preserve the density will survive -- but matter which is invariant
under density preserving diffeomorphisms is an ideal liquid i.e.\ a
superfluid.

\subsection{Superconductivity}

As in the one-dimensional case, it is almost obvious that a dilute
incommensurate system of solitons forms a compressible liquid.
Indeed, a soliton position is not fixed relatively to the rest of the
system.  It can be translated to any place in the system without
changing the energy -- zero modes are translationally degenerate.
Therefore, if a soliton moves slowly, say with a momentum $\vec k$,
the energy of the system changes by $vk$.  Since the interaction
between solitons is short range, the same remains true for dilute
system of solitons \footnote{ At larger density solitons may form a
Wigner crystal, i.e.\ a solid, or the interaction may destroy a bound
state between a soliton and electrons (whatever comes first).  It may
happen that the soliton density is commensurate with the lattice.  In
this case solitons form a solid and are pinned by the lattice.  If
commensurability is weak, the solid may be melted by quantum
fluctuations.  We do not discuss these questions here.} Thus we assume
the solitons form a liquid.  Each soliton carries a density of charge
carriers and an electric charge according to its topological charge
(\ref{t1}).  Translation degeneracy of the zero modes therefore
implies longitudinal sound in modulation of electronic density and
this conforms to the first part of the Landau criterion.

Furthermore, the narrow band formed by electrons trapped in solitons
is always filled and separated from the rest of the spectrum.
Therefore all excitations, except coherent modulation of charge
density and topological density, cost non-zero energy.  Indeed, one
particle excitations consist of a particle being taken away from the
soliton core.  Since all particles form bound states with solitons,
this costs energy i.e.\ the excitation has a gap.  This conforms to
the second part of the Landau criterion.  To summarize, we conclude
that the Landau criterion is satisfied and the topological liquid is a
superfluid.

To formalize these arguments let us consider for example the basic
model (\ref{m1}) with
\begin{equation}\label{ss1}
H=\sum_{\sigma=1,2}\psi^{\dag}_\sigma[\alpha_\mu(i\partial_\mu+a_\mu)
+\beta m]\psi_\sigma
\end{equation}
in two dimensions.  At low energy the system develops a density of
magnetic flux determined by the density of extra particles:
\begin{equation}
\label{ss2} \rho(x)=-{1\over \pi}\vec\nabla \times\vec a=-{1\over
\pi}F(x) \enspace .
\end{equation}
When the solitons move, they are followed by particles, implying
\begin{equation}
\vec j^*(x)=-{1\over \pi}(\partial_t\vec a-\vec\nabla a_0)
= -{1\over \pi}\vec E \en ,
\end{equation}

where $j^*_i=\varepsilon_{ik}j_k$, and an additional factor 2 comes
from the spin.  A homogeneous density $\bar\rho$ gives rise to a
homogeneous magnetic field $\langle F\rangle$ and no electric field.
This is the first level of adiabatic approximation.  To study
modulation of the density and to see that there is no dissipation, we
want to see how the gauge field fluctuates.  Its fluctuations are
determined by the motion of electrons in the presence of magnetic
flux $F/\pi=\bar\rho$.  The relation between the density and topology
implies that the first two Landau levels are completely filled and in
addition they separated by the gap $m$ from the upper Dirac band.  In
other words, all single particle excitations have a gap, which means
that
excitations over the gap produce an energy ${1\over 2}\varepsilon
\vec E^2+{1\over {2\chi}}(F-{\langle F\rangle})^2$.  Here
$\varepsilon $ and $\chi$ are the dielectric constant and diamagnetic
susceptibility of electronic zero modes, respectively.  As a result
static
fluctuations of the topological charge $F(x)-\langle F\rangle$ and
topological current $\vec E(x)$ are short range:

\begin{equation}
\langle (F(x)-{\langle F\rangle}),(F(x')-{\langle F\rangle})\rangle
=\chi \delta(x-x') \en ,
\end{equation}
\begin{equation}
\langle\vec E(x),\vec E(x')\rangle={\varepsilon}^{-1}
\delta(x-x')\enspace .
\end{equation}

We can now obtain density and current correlation functions since they
are directly related to fluctuations of the topological charge.
Combining them with the continuity condition, we obtain the linear
response function for a superfluid with $\lambda^{-2}=\chi$ and
$v^2=\varepsilon\chi$ in (\ref{L1}) and (\ref{L2}).

A similar strategy applies for the 2D model (\ref{s5}).  Analysis of
the three dimensional model is more complicated and involves some new
features \cite{Wiegmann}.  The physical picture, however, remains the
same, as briefly indicated below.

Since all single particle excitations have a gap, we expect that all
low energy excitations are fluctuations of the topological charge
$q_0$ and topological current $q_i$. For the same reason, we know
that static fluctuations are short range.  We also know that there is
a longitudinal sound since the topological configuration possesses a
soft translational mode.  Assuming that solitons do not form a
crystal, we must conclude that the low energy topological excitations
obey hydrodynamics:
\begin{equation}\label{q}
H={\nu^2\over 2\kappa}[q_0^2+v_0^{-2}{\vec q}^2] \en .
\end{equation}
where $(q_0,\vec q)$ is a density of the topological charge and topological
current of  solitons,

Finally, as a result of the topological instability, the density of
electric charge (current) is directly related to the topological
charge (current)
\be
j_\mu=\nu q_\mu \en ,
\ee
which, together with (\ref{q}), gives the hydrodynamics of the ideal
charged liquid,
$$H={1\over 2\kappa}[(\rho-\rho_0)^2+v_0^{-2}{\vec j}^2] $$.
i.e.\ hydrodynamics of the superconductor (\ref{L6}).

\subsection{Topological Superconductivity and Bosonization in Higher
Dimensions}

Bosonization is a fascinating phenomenon of one-dimensional
physics -- collective modes of interactive fermionic systems may be
described by bosonic particles.  Moreover, fermions themselves can be
treated as solitons of bosonic non-linear waves.  The physical reasons
behind bosonization are the same as for the Frohlich
superconductivity.
In fact, a generic one-dimensional incommensurate system develops some
density wave (CDW,SDW, etc.) as a low energy excitation.  These are
bosonized fermionic collective modes with
\begin{equation}
H={\bar\rho\over 2}[\pi^2+v_0^2(\partial_x \varphi)^2] \en .
\end{equation}
(Here $\pi$ is the canonical momentum of the bosonic field $\varphi$.)
In other words, bosonization in one dimension implies hydrodynamics.
If the one-dimensional system is commensurate, the Hamiltonian of the density
wave acquires some non-linear terms, as in the Sine-Gordon model, for
example.

This interpretation allows the extension of bosonization into higher
dimensions.  We shall say that a fermionic system can be bosonized if
its low energy excitations can be described entirely in terms of
currents, such as when an effective Hamiltonian of low energy dynamics
has the Sugawara form, i.e. may be written as a bilinear form of
currents.  Recall that this form implies the hydrodynamics of the
superfluid (6).  The Sugawara form is a linear form of quantum
hydrodynamics \cite{Landau}:
\begin{equation}
H={1\over 2}\rho (x) { \vec v(y)}^2,~~~~~~~~[\rho (x),\vec
v(y)]=i\vec\nabla\delta (x-y),~~~~~~\vec j=\rho \vec v \en .
\end{equation}

The bosonized version of linear hydrodynamics appears in terms of
a displacement $\vec u$.  Let us solve the continuity condition by
setting $\rho-\bar\rho=\bar\rho\nabla\vec u $ and $\vec
j=\bar\rho\partial_t\vec u$.  Then (6) becomes
\be
L={\bar\rho\over 2}[(\partial_t\vec u)^2-v_0^2
(\nabla\vec u)^2] \en .
\ee

In contrast to one dimension, not every two-or three-dimensional
fermionic system can be bosonized, but only those which are
superfluids.  For instance a fundamental property of the Fermi liquid
is that it cannot be bosonized.  Dissipation and gapless excitons are
its own characteristic features.

We have indicated that a fermionic theory with a topological charge in
the ground state can be bosonized, regardless of its dimension.  From
this point of view  bosonization can be treated as a projection
onto a zero mode narrow band.

\section {A Gauge Model for Correlated Electronic Systems}\label{V}

Strongly correlated electronic systems represent physical systems
where topological fluids may appear.  In this section we sketch the
derivation of a basic model of a topological fluid from the Hubbard
model or the ``t - J'' model -- a canonical model of strongly
correlated electronic systems.  We also emphasize a set of assumptions
and approximations which are necessary to separate the physics of
topological fluids from a variety of other physical phenomena which
occur in a correlated electronic system.

\subsection{A Model for the Doped Mott Insulator -- Resonant Valence Bonds
and Chirality}
The ``$t - J$'' model is defined by
\begin{equation}\label{g1}
H = \sum\nolimits\limits_{ ij}^{} { t}_{ij} { c}_{i\sigma}^+
{ c}_{j\sigma} +{ J}_{ij} {\vec{S}}_{ i} {\vec{S}}_{j} \quad
({\rm all}\; {J}_{i,j}>0)\en ,
\end{equation}
where ${\vec{S}}_{ i} = { c_{i\sigma}}{\vec{\sigma}_{ \sigma\sigma '}
{c}_{i\sigma '}}$ is a spin operator of an electron at the lattice
site $ i$. A hopping amplitude $t_{ij}$ and an exchange amplitude
$J_{ij}$ connect the nearest sites and the total number of electrons
is close to the number of lattice sites: ${ N}_{e} = { N}_{0} (1 -
{\bar\rho})$, while a strong Coulomb interaction does not allow doubly
occupied states:
\begin{equation}\label{g2}
{n}_{i} =\sum_{\sigma =\uparrow,\downarrow} {c}_{i\sigma}^{+}
{c}_{i\sigma}^{\phantom +}= 0\quad\rm or\quad 1  \enspace .
\end{equation}
At zero doping the model describes the Heisenberg antiferromagnet
\begin{equation}
H =\sum_{ab} { J}_{ab} {\vec{S}}_{ a}{\vec{S}}_{b} \en ,
\end{equation}
with spins ${\vec{S}_{a}}$ and ${\vec{S}_{b}}$ on sublattices A and B,
respectively.

In addition to an average value of spin (classical description) there
are two operators which characterize the ground state of an
antiferromagnet, namely {\em density of energy}
\begin{equation}
{\varepsilon }_{ij} =({1 \over 4}+{\vec{S}}_{i}{\vec{S}}_{j})
\end{equation}
and {\em chirality} or measure of topological
order \cite{Scripta,WWZ,KW}
\begin{equation}
W(C) = {\rm tr} \prod_{i\epsilon C}  ({1 / 2}+\vec{\sigma}{\vec{S}_{i}})
\en ,
\end{equation}
where $\vec{\sigma}$ are Pauli matrices, and $C$ is a lattice
contour.

The latter operator is of particular importantance for the doped case
since it determines the correlation of electronic phases at different
spatial points.

These two operators are related to the amplitude and phase
$\Delta_{ij}$ of Anderson's Resonance Valence Bond (RVB) through
\begin{equation}
\varepsilon_{ij}  =\left|{{ \Delta }_{ij}}\right|{}^{2}
\end{equation}
and
\begin{equation}
\label{chi}  W (C) = \prod_{C} {\Delta }_{ij} \en .
\end{equation}
It follows from this definition that the RVB is a gauge field.   It
can be locally transformed by a U(1) transformation
\begin{equation}
{\Delta }_{ij} \rightarrow {\Delta }_{ij}{e}^{i(\alpha_i-\alpha _j) }
\end{equation}
and by a symmetry group  of the crystal class
\begin{equation}
\Delta  \rightarrow {C}^{-1} \Delta\null C   \en ,
\end{equation}
without changing the physical properties.   In terms of this field,
the topological
order parameter $W(C)$ acquires the form of a lattice Wilson loop:
\begin{equation}
W (C) = {e}^{i\phi  ( C )} \en .
\end{equation}
Roughly speaking, the flux of the RVB field
\begin{equation}
{e}^{i\phi  ( C )}=\prod_{ C}  e^{iA_{ij}} \en ,
\end{equation}
where $A_{ij}$ is a phase of $\Delta_{ij}$, represents a ``magnetic''
flux, and penetrates through a surface enclosed by the contour $C$.
The RVB field can be formally introduced by fermionic representation
of a spin operator:
\begin{equation}
\vec{ S}={c}_{if}^{+}{\vec{\sigma }}_{\sigma\sigma '}{c}_{i\sigma '}
\end{equation}
with the condition
\begin{equation}\label{constr}
\sum\nolimits\limits_{\sigma}^{} { c}_{i\sigma}^{+}{ c}_{i\sigma} =1
\end{equation}
so that
\begin{equation}
{\Delta }_{ij} =\sum_{ \sigma}  { c}_{i\sigma}^{+}{ c}_{j\sigma}
\en .
\end{equation}

Using this representation the Heisenberg Hamiltonian can be explicitly written
in terms of the RVB field \cite{BA,IL} as

\begin{equation}
H _A =\sum_{ \sigma ,\langle ab \rangle} {c}_{\sigma i}^{ +}
{ U}_{ij }{ c}_{\sigma j} +{ A}_{ 0}( i)({ c}_{\sigma i}^{ +}
{ c}_{\sigma i }-1)+{U}_{ij} { J}_{jk}^{-1}{ U}_{ki}
\en .
\end{equation}
The field ${A_0}$ is a Lagrange multiplier which ensures the
constraint (\ref{g2}), while the phase of the conjugated RVB field
$U_{ij}$ forms a time and space component of a gauge field.  The
constraint (\ref{constr}) is relaxed after doping, when a small
density of empty sites appear on the lattice,
\begin{equation}
\sum_{\sigma} { c}_{\sigma i}^{ +}{c}_{\sigma_ i}=1- h_{i}^{+} h_{i}
\en ,
\end{equation}
where ${h_i}$ is a dopant (hole) operator.  An electron operator is
then represented by the product of a spinon and a holon, $ c_{\sigma
i}h^{\dag}_i$.  The hopping Hamiltonian is therefore
\begin{equation}
H=t\sum_{\langle\vecc a,\vecc b\rangle}h^{\dag}(\vecc a)
h(\vecc b)\Delta_{\vecc a,\vecc b}+\;\;{\rm c.c.}
\end{equation}
where $\vecc a$ and $\vecc b$ are sites of sublattices $A$ and $B$ and
$\Delta_{\vecc a,\vecc b}= \langle\sum_{\sigma}
c^{\dag}_{\sigma}(\vecc a)c_{\sigma}(\vecc b)\rangle$.  So far all
manipulations were exact.

Depending on parameters this model may describe physically different
situations.  We are interested in a situation when fluctuations of the
bond energy (modulus of $\Delta_{i,j}$) are much less and much slower
that fluctuations of the bond phase (chirality).  If their scales are
separated we refer to this state as quantum antiferromagnetism.  Let
us stress that in the Ne\'el state the modulus and phase of bonds
fluctuate similarly and cannot be separated.  In this case holes form
the Fermi-liquid and interact weakly with spin fluctuations.

\subsubsection{Adiabatic Approximation.}

Some progress can be made if the magnetic background can be treated
adiabatically.  This means that a hole moves in a slowly varying spin
configuration, similar to the Peierls-Frohlich problem where electrons
move with a slowly varying distortion.  In this case we may use a
semi-classical strategy: first find a static $\bar\Delta$ and $ \bar
A_0$ which minimize the energy of the system with a given doping,
\begin{equation}\label{g3}
{\Delta }_{ij} ={ \delta  \langle H \rangle \over  \delta {U}_{ij}}
\en ,
\end{equation}
and then take into account quantum fluctuations around the static
mean field.

To apply the semi-classical strategy one more step is required.  The
point is that even if the hopping amplitude is bigger than the
exchange energy, the dynamics of spins cannot be considered
adiabatically -- in the antiferromagnet each jump of a hole sharply
changes the spin configuration by flipping a spin on a sublattice.
However, two consecutive jumps bring a hole to the same sublattice, so
the spin configuration remains approximately unchanged.  Thus, to
apply adiabatic arguments we must first integrate over fast and sharp
processes.  Perhaps, the easiest way to implement the adiabatic
approximation is to introduce a difference between energies of a hole
on different sublattices by adding the term $\mu (\sum_{\vecc
a}c_{\vecc a}^{\dag}c_{\vecc a}-\sum_{\vecc b}c_{\vecc
b}^{\dag}c_{\vecc b})$ to the Hamiltonian.  If the hopping energy $t$
is less than $\mu$ ( adiabatic parameter), a holon and a doublon
appear on different sublattices only virtually.  Therefore in leading
order in $t/\mu$ one may consider only processes of two consecutive
hoppings described by the effective holon Hamiltonian
\begin{equation}
H=t^{\prime}\Big(\sum_{\vecc a,\vecc a^{\prime}}h^{\dag}(\vecc a)
\Delta_{\vecc a,\vecc a^{\prime}}^2h(\vecc a^{\prime})+
\sum_{\vecc b,\vecc b^{\prime}}{h^{\dag}(\vecc b)}
\Delta_{\vecc b,\vecc b^{\prime}}^2h(\vecc b^{\prime})+{\rm c.c.}\Big)
\end{equation}
 where
\begin{equation} \Delta_{\vecc a,\vecc
a^{\prime}}^2=\sum_{\vecc b^{\prime}}\Delta_{\vecc a,\vecc b^{\prime}}
\Delta^{*}_{\vecc b^{\prime},\vecc a^{\prime}}~~~~~
\Delta_{\vecc b,\vecc
a^{\prime}}^2=\sum_{\vecc a^{\prime}}\Delta_{\vecc b,\vecc a^{\prime}}
\Delta^{*}_{\vecc a^{\prime},\vecc b^{\prime}}
\end{equation}
are hopping operators within the same sublattice.
After all, we suggest that this Hamiltonian be considered as
phenomenological.

Now as long as $t^{\prime}\sim t^2 /\mu >>J$, one can treat magnetic
fluctuations adiabatically.  A solution of the mean field equation
(\ref{g3}) depends on a particular exchange coupling.  Here we are
interested in a solution where $|\Delta_{ab}|$ slightly varies from
bond to bond, but never vanishes: $\Delta_{ab}\neq 0$.  Then the mean
field values of $\Delta_{ij}$ and $U_{ij}$ are proportional to each
other.  After all these considerations we end up with the basic
Hamiltonian
\begin{eqnarray}\label{M1}
H & = & \tilde{t}
\Big(\sum_{\vecc a,\vecc a^{\prime}}h^{\dag}(\vecc
a)\Delta_{\vecc a,\vecc a^{\prime}}^2 h(\vecc a^{\prime})
+ \sum_{\vecc b,\vecc
b^{\prime}}h^{\dag}(\vecc b)\Delta_{\vecc b,\vecc b^{\prime}}^2 h(\vecc
b^{\prime})+{\rm c.c.}\Big)+ \tilde{J}\sum_{\vecc a,\vecc b}
c^{\dag }_{\sigma,\vecc a} {\Delta}_{\vecc a } c_{\sigma,\vecc b}
\nonumber\\[1mm]
& & + \sum_i A_0 ( c_{\sigma i}^{ \dag} c_{\sigma i }+h_i^{\dag}
h_i-1) \en ,
\end{eqnarray}
where $\tilde{t}\sim t'J^{-1}$ and $\tilde{J}\sim J$ are
phenomenological constants and we set $|\Delta_{ij}|=1$.  To remain
within adiabatic approximation we also assume that
$\tilde{t}>\tilde{J}$.

We refer this model as the gauge model of correlated electronic
systems.  In addition to holons and spinons it has a gauge field,
the phase of the bond $\Delta_{ij}=\exp{iA_{ij}}$, and a Lagrange
multiplier $A_0$ as a dynamical gauge field.  Let us also notice
that the gauge field has no energy of its own.

The hopping part of the model has an important symmetry which reflects
the symmetry of the square lattice.  The spectrum of the Hamiltonian
is symmetric, namely, if there is an eigenstate with an energy $E$,
then there is an eigenstate with the energy $-E$.  To see this one
must
note that the Hamiltonian (\ref{M1}) changes sign if one replaces
$c_{\vecc b}$ and $h_{\vecc b}$ to $-c_{\vecc b}$ and $-h_{\vecc b}$
and keeps $c_{\vecc a}$ and $h_{\vecc a}$ unchanged.  It also means
that there is a matrix $\Gamma$ which anticommutes with the
Hamiltonian.  In the sublattice basis it is a diagonal matrix
\begin{equation}\label{Gamma}
\Gamma={\rm diag}(1,-1) \en.
\end{equation}

\subsection{Flux Phase}

The next step in the adiabatic approximation is to determine a mean
value of the gauge field $A_{ij},A_0$. We approach the problem in
two stages.  At first, we neglect a small doping and find the most
favorable configuration of $U_{ij}$ for the half-filled case.  Then
we discuss the effect of doping.  The first stage is again approached
in two parts.  Assuming that $|\Delta_{ab}|$ is a constant, we first
find the phase of $\Delta_{ab}$ or $U_{ab}$ i.e., the ground state
value of the chirality (\ref{chi}) and later consider a small
variation of $|\Delta_{ab}|$.

At the first step, we suppose that a variation of $|\Delta_{ab}|$ in
space is small and may be found on the basis of continuous theory.
Let us therefore set $|U_{ab}| = \Delta_0 J_{ab} = t_s$ in the
first approximation.  Then the only degree of freedom to be determined
is the chirality $W ( C )={ \Delta }_{ 0}^{ L (C)}\exp (i \phi(C))$.
It has been suggested in Refs.\cite{Lederer,AM,Kotl} and
has been proved recently by E.Lieb \cite{Lieb} that the flux $\pi$ per
elementary plaquette provides the minimum of the energy:
\begin{equation}
W(P) = \Delta_0^4 (-1) \en .
\end{equation}
This state is known as a flux phase or chiral phase.

\subsection{Continuum Limit}

As usual, in order to go to the continuum limit we must separate fast
and slow variables.  The fast variables are represented by the gauge
potential of the background flux.  In the flux state translations do
not commute any longer.  They form a group of magnetic translations:
\begin{equation}\label{c1}
T_{\vecc e_i}T_{\vecc e_j}+T_{\vecc
e_j}T_{\vecc e_i}=0 \en ,
\end{equation}
where $\vecc e_i$ are
primitive lattice vectors.  Apart from spinor representations of
SO$_3$, it is also reprented by $2^d$ matrices of dimension
$d\times{d}$ (where $d$ is a spatial dimension) which obey the
Clifford algebra
\begin{equation}
\{{\alpha}_i, {\alpha}_j\} = 2\delta_{ij} \en .
\end{equation}

Setting $\Delta_{ab}$ to its mean field value $\overline{\Delta}_{ab}
= {\alpha}_\mu$, and $\overline{A}_0 (a) = \overline{A}_0 (b) = m$, we
get the nearest neighbour part of the Hamiltonian in the form
\begin{equation}
H =t\sum_{ i =1}^d  {
T}_{\vecc e_i}{ e}^{ 2i (\vecc{k}{\vecc{e}}_{i})}+ m\beta
=t \sum_{i=1}^{d}\alpha_{i}\cos 2{k}_{i} + m\beta
\end{equation}
where the wave vector $\vec k=( k_x,k_y,...)$ belongs to the reduced
Brillouin zone $-{\pi\over 2} < k_i < {\pi \over 2 }$, while the
matrix
$\beta = \pm 1$ depending on whether a site belongs to sublattice $A $
or $ B$, and $\{ {\alpha}_i,{\beta} \}=0$.  To find the spectrum it
is sufficient to consider the square of $H$.  Due to (\ref{c1}) one
has
\begin{equation} H^2 =
\sum_{i = 1}^{d} {\cos{2k_i}}^2 + m^2 \en .
\end{equation}

The spectrum contains two Dirac bands which touch at $k_{\rm
F}=({\pi\over {2}},{\pi\over {2}},...)$.  The lower band is completely
filled.  Expanding in small momentum $\vec {p} = \vec {k} - \vec
{k}_F$ around the Dirac point, we get the Dirac Hamiltonian
\begin{equation}\label{Dirac}
H =\sum_\sigma \psi_\sigma^{\dagger}\left(\vec{\alpha}\vec{\partial} +
m\right) \psi_\sigma
\end{equation}
where the Dirac spinor $\psi_\sigma$ is the slowly varying part of
$c_\sigma$.  To see this, let us denote a lattice site $\vecc R$ as
\begin{equation}
\vecc{ R}=\sum\nolimits\limits_{i=1}^{3} (2{x}_{i}+(r_{i}-1))
{\vec{e}}_{i} \en ,
\end{equation}
where $r_i = 1,2$ and $x_i$ are integers.   If $r_i = 1$, the site
belongs to the sublattice $ A$; if $r_i = 2$ then to a sublattice
$B$.   Then
\begin{equation}
{ \psi }_{{\vecc r},\sigma}(\vecc{x})\approx\exp(i{\vecc{k}}_{\sigma}
\vecc{R}){c}_{\vecc{R},\sigma} \en .
\end{equation}

Let us mention that the dimension of the matrices $\alpha$ is twice
that of the irreducible representation of the Clifford algebra, so
they belong to a reducible representation.  This is the well known
lattice doubling -- a lattice always produces an even number of
fermionic species.  To summarize, let us denote ~$ T_{\vec e_i} =
\alpha_i;~ \gamma_0 = \beta;~ \gamma_i = { \beta} { \alpha}_i$.  Then
$ \big\{ { \alpha}_i,~{ \alpha}_j \big \} = 2 \delta_{ij};~ { \beta} {
\alpha}_i~ { \beta} = { \alpha}_i;~ { \beta}^2 = 1$ and the
$\gamma_\mu$'s form an algebra of Dirac matrices, $\left \{
\gamma_\mu~ \gamma_\nu \right \} = 2g_{\mu \nu}~$ .

This type of mean field theory is essentially invalid, because it does
not manifestly reflect the main physical property of the theory -- its
gauge invariance.  However, the gauge symmetry is restored by
fluctuations.  Before we discuss gauge fluctuations around the mean
field, let us choose a convenient basis of the $\alpha$ matrices on a
square lattice.

\subsection {Lattice adopted basis}

Expansion around the mean field value follows the strategy developed
in polyacetylene physics \cite{Heeger,BrKr} -- only processes with a
momentum transfer of $2k_{\rm F}$ are relevant.  Others lead to terms
with higher order derivatives.  For the sake of simplicity we consider
here the basis adopted on the 2D square at half filling.
Generalization to the three-dimensional cubic lattice is developed in
Ref.\cite{Wiegmann}.  Since at half filling the period is doubled, we
must consider the four nearest points of a lattice sell as a set of
discrete variables.  Let us denote them in terms of Cartesian
coordinates: $\vecc r = \left( r_x, r_y \right): r_i = 1,~ 2.$ Then
${\vec \alpha}$ can be chosen in the form
\begin{equation}
(\alpha_x)_{\vecc r \vecc r^\prime}  =  (\sigma_1)_{r_x
r^\prime_x} \otimes  (\sigma_3)_{r_y~r^\prime_y}  \en ,
\end{equation}
\begin{equation}
\alpha_y =  \font\elevenrm=cmr10 scaled 1095
\hbox{\hbox{1}\hskip-2.5pt\lower.1pt\hbox{\elevenrm l}}
 \otimes \sigma_1~ ;~~
\beta = \sigma_1 \otimes \sigma_2
\hbox{\hbox{1}\hskip-2.5pt\lower.1pt\hbox{\elevenrm l}} \en ,
\end{equation}
where the $\sigma$'s are Pauli matrices.  In addition there are
three more matrices $\tau_x,\tau_y $ and $\tau_z$ which commute with
the matrices $\alpha$.  We choose them to form a basis of SU(2),
$[\tau^{i},\tau^{j}]=i\epsilon_{ijk} \tau^{k}$:
\begin{equation}
 \tau_1 = \sigma_1 \font\elevenrm=cmr10 scaled 1095
\hbox{\hbox{1}\hskip-2.5pt\lower.1pt\hbox{\elevenrm l}}~;~~
\tau_2 = \sigma_3  \otimes \sigma_1~;~~
\tau_3 = \sigma_2 \otimes \sigma_1 \en .
\end{equation}
There is of course a direct product basis in which spin and isospin
appear as in independent spaces:
\begin{equation}
{ {\vec \alpha}} \to M { {\vec \alpha}} M^{-1} =
\font\elevenrm=cmr10 scaled 1095
\hbox{\hbox{1}\hskip-2.5pt\lower.1pt\hbox{\elevenrm l}}
\otimes \alpha_i~;~~ { { \beta}} \to
\font\elevenrm=cmr10 scaled 1095
\hbox{\hbox{1}\hskip-2.5pt\lower.1pt\hbox{\elevenrm l}}
\otimes \beta~;~~{ {\vec \tau}} \to {\vec \tau} \otimes
\font\elevenrm=cmr10 scaled 1095
\hbox{\hbox{1}\hskip-2.5pt\lower.1pt\hbox{\elevenrm l}} \en .
\end{equation}

In continuous theory, the amplitude $\Delta_{ab}$ fluctuates around
its mean field value such that these fluctuations are smooth, but not
necessarily small.  Fluctuations of the modulus of $\Delta_{ab}$ are
small and we neglect them, but fluctuations of the phase of
$\Delta_{ab}$ restore the gauge invariance and may not be small.  To
describe them we write
\be\Delta_{ab}=e^{\Lambda_i}\alpha_i T_i \en ,
\ee
where $T_i=\exp({-\partial_i})$ is a translation operator and
$\Lambda_i$ a diagonal matrix. (Fluctuations may change the value of a
bond, but do not introduce a new bond.)  Four diagonal matrices which
form a basis for fluctuations may be written in terms of the Dirac
basis.  They are $ \font\elevenrm=cmr10 scaled 1095
\hbox{\hbox{1}\hskip-2.5pt\lower.1pt\hbox{\elevenrm l}}\otimes
\font\elevenrm=cmr10 scaled 1095
\hbox{\hbox{1}\hskip-2.5pt\lower.1pt\hbox{\elevenrm l}}$ and
given by
\begin{eqnarray}
\alpha_1\tau^1 & = & \font\elevenrm=cmr10 scaled 1095
\hbox{\hbox{1}\hskip-2.5pt\lower.1pt\hbox{\elevenrm l}} \otimes  \sigma^3
\en ,\nonumber \\
\alpha_2\tau^2 & = &  \sigma^3  \otimes  \font\elevenrm=cmr10 scaled 1095
\hbox{\hbox{1}\hskip-2.5pt\lower.1pt\hbox{\elevenrm l}}
\en ,\nonumber \\
\beta\tau^3    & = & \sigma^3  \otimes  \sigma^3 \en .
\end{eqnarray}

The simplest way to include fluctuations of the phase is to restore
the gauge invariance.  The original lattice model is invariant under
the $U(1)$ gauge transformation $c_i\rightarrow c_i\exp{i\varphi_i}$.
In the continuum limit where each of the four sites is treated
separately, the gauge group becomes a four parameter group, and
accordingly the phase of the gauge transformation is an arbitrary
diagonal matrix:
\begin{eqnarray}\label{w1}
\psi & \rightarrow & e^{i\varphi_1}\psi                 \nonumber \\
\psi & \rightarrow & e^{i\varphi_2\alpha_1\tau^1}\psi   \nonumber \\
\psi & \rightarrow & e^{i\varphi_3\alpha_2\tau^2}\psi   \nonumber \\
\psi & \rightarrow & e^{i\varphi_4\beta\tau^3}\psi \en .
\end{eqnarray}

To restore the gauge invariance of the mean field Hamiltonian
(\ref{Dirac}) we must add compensating fields:
\begin{eqnarray}\label{E1}
 H&=&
\exp[iE_1\beta\tau^3+iG_1\alpha_1\tau^1+
iK_1\alpha_2\tau^2]\alpha_{1}(i\partial_{1}+a_1)\exp[-iE_1\beta\tau^3]
\nonumber \\
&+&\exp[iE_2\beta\tau^3+iG_2\alpha_1\tau^1+
iK_2\alpha_2\tau^2]\alpha_{2}(i\partial_{2}+a_2)\exp[-iE_2\beta\tau^3]
\nonumber \\
&+&  \beta\tau^3 m \en .
\end{eqnarray}
These compensating fields are gauge fields.  Under the transformation
(\ref{w1}) they transforms as follows:
\begin{eqnarray}
A_\mu & \rightarrow & A_\mu+\partial_\mu\varphi_1        \nonumber \\
G_\mu & \rightarrow & G_\mu+\varphi_2  \nonumber \\
K_\mu & \rightarrow & K_\mu+\varphi_3\nonumber \\
E_\mu & \rightarrow & E_\mu+\varphi_4
\end{eqnarray}

The quantities $E=E_1-E_2, G=G_1-G_2, K=K_1-K_2$ and
$F=\vec\nabla\times\vec a$ are therefore gauge invariant objects.
They are not small, so the exponents in (\ref{E1}) cannot be expanded.
The four types of ``magnetic field'' describe fluctuations of the flux
in four adjacent plaquettes and implement the symmetry of the square
lattice.

Fixing a gauge, one may slightly simplify the Hamiltonian (\ref{E1}).
One may choose $G_2=0,K_1=0,E_2=0$, in which case the Hamiltonian has
the form
\begin{eqnarray}\label{E}
 H& = &
\exp[iE\beta\tau^3]\alpha_{1}(i\partial_{1}+a_1)\exp[-iE\beta\tau^3]
\alpha_{2}(i\partial_{2}+a_2)\exp[-iE_2\beta\tau^3] \nonumber \\
& & \qquad
+\partial_{1}G\tau^1+\partial_{2}K\tau^2
+\beta\tau^3 m
\end{eqnarray}
We note that even though the terms $i\partial_{1}G$ and
$\partial_{2}K$ appear to be temporal components of the SU(2) gauge
field, their symmetry is different.  They are components of the spinor
connection of the original crystal group.

Fluctuations of the modulus can also be obtained from this symmetry.
First of all the Hamiltonian must anticommute with the matrix
(\ref{Gamma}), which in our basis is $\Gamma=\beta\tau^3$.  Secondly,
since the Hamiltonian connects only the nearest lattice sites, it must
be a tensor product of diagonal and anti-diagonal matrices in the
sublattice basis.  Except for the matrices $\alpha_i, \alpha_i\tau^3,
\tau^1, \tau_2$ which already appeared in the Hamiltonian, three more
such matrices, $\beta\alpha_i\tau^3, \beta\tau^1, \beta\tau^2$,
describe fluctuations of the modulus of $\Delta_{ab}$.  Taking them
into account too, finally gives
\begin{eqnarray}\label{E3}
 H&=&
\exp[iE\beta\tau^3]\alpha_{1}(i\partial_{1}+a_1)\exp[-iE\beta\tau^3]
\alpha_{2}(i\partial_{2}+a_2)\nonumber \\
&&+\partial_{1}G\tau^1+\partial_{2}K\tau^2+i\beta\alpha_i\tau^3
\Phi_i+\beta\tau^a\Phi^a +\beta\tau^3 m
\end{eqnarray}
where $\Phi_i$, $\Phi^a$ are components of the modulus of $\Delta$ in a
proper basis.
As a result,  the basic model for the doped Mott insulator in the flux
phase has a form:
\begin{equation}\label{E4}
{\cal H}={1\over {2m}}h^{\dag}(x) H^2h(x)+ \psi^{\dag}(x) H\psi(x)+
{1\over {2\lambda}}(\Phi_i^2+(\Phi^a)^2) \en .
\end{equation}

For further explicit reference, let us present the Hamiltonian $H$ for
the three dimensional flux phase.  In this case translations are
$8\times 8$ matrices.  They can be chosen in the form
\begin{eqnarray}
T_{1}&=&\alpha_{1}=\sigma_{1}\otimes\sigma_{3}\otimes{\bf 1}
\nonumber\\
T_{2}&=&\alpha_{2}={\bf 1}\otimes\sigma_{1}\otimes\sigma_{3}
\nonumber\\
T_{3}&=&\alpha_{3}=\sigma_{3}\otimes{\bf 1}\otimes\sigma_{1} \en .
\end{eqnarray}
The fourth Dirac matrix $\beta$ may be chosen to be diagonal,
\be
\beta=\sigma_{3}\otimes\sigma_{3}\otimes{\sigma_{3}} \en .
\ee
Four other matrices which commute with the Dirac matrices are
translations along face diagonals,
\begin{eqnarray}
T_{12}&=&\tau^{3}=\sigma_{2}\otimes\sigma_{1}\otimes{\sigma_{3}}
\nonumber \\
T_{23}&=&\tau^{1}=\sigma_{3}\otimes\sigma_{2}\otimes{\sigma_{1}}
\nonumber \\
T_{31}&=&\tau^{2}=\sigma_{1}\otimes\sigma_{3}\otimes{\sigma_{2}}
\en,
\end{eqnarray}
and translation along the space diagonal,
\begin{eqnarray}
T_{123}=i\gamma_{5}\gamma_{0}=
\gamma_{1}\gamma_{2}\gamma_{3}=\alpha_{1}\alpha_{2}\alpha_{3}\beta
=\sigma_{1}\otimes\sigma_{1}\otimes{\sigma_{1}} \en .
\end{eqnarray}
This last operator is known as the parity operator.

The  8 diagonal matrices which describe fluctuations are
\begin{eqnarray}
&O_{1}={\bf 1}\otimes{\bf 1}\otimes{\bf 1}&\quad
O_{2}=\sigma_{3}\otimes{\sigma_{3}}\otimes\sigma_{3}\nonumber\\
&O_{3}=\sigma_{3}\otimes{\bf 1}\otimes{\bf 1}&\quad
O_{4}={\bf 1}\otimes\sigma_{3}\otimes{\bf 1}\nonumber\\
&O_{5}={\bf 1}\otimes{\bf 1}\otimes\sigma_{3}&\quad
O_{6}=\sigma_{3}\otimes\sigma_{3}\otimes{\bf 1}\nonumber\\
&O_{7}=\sigma_{3}\otimes{\bf 1}\otimes\sigma_{3}&\quad
O_{8}={\bf 1}\otimes\sigma_{3}\otimes\sigma_{3} \en ,
\end{eqnarray}
%HBG? Paul, are the 6 matrices below supposed to be related to the 8
%above? If so, what is the relation and why the forms below?
%
%They are
%\be
%\tau^{1}\alpha_{2}\alpha_{3}\beta,~
%\tau^{2}\alpha_{1}\alpha_{3}\beta,~
%\tau^{3}\alpha_{1}\alpha_{2}\beta,~
%\tau^{3}\alpha_{1}\alpha_{2},~
%\tau^{2}\alpha_{1}\alpha_{3},~
%\tau^{1}\alpha_{2}\alpha_{3}
%\ee
Following the procedure we discussed for the 2D case, one obtains
\begin{eqnarray}\label{E5}
H &=& \sum_i\exp\Big[iE_i\beta+\sum_{k\ne l\ne m\ne
n}(iG_i^k\tau^{l}\alpha_{m}\alpha_{n}\beta+iK_i^k\tau^{l}\alpha_{m}
\alpha_{n})\Big]\alpha_{i}(i\partial_{i}+a_i)\times \nonumber \\
&&\times\exp\Big[iE_i\beta+\sum_{k\ne l\ne m\ne n}
(iG_i^k\tau^{l}\alpha_{m}\alpha_{n}\beta+
iK_i^k\tau^{l}\alpha_{m}\alpha_{n})\Big]\nonumber\\
&&+i\alpha_i\beta \Phi_i+i\alpha_i\beta\tau_i \Phi_i^j+
\alpha_1\alpha_2\alpha_3\tau_i\Phi^i
\end{eqnarray}

Despite some efforts, topology of the numerous fields appeared in
the models (\ref{E4},\ref{E5}) and their  anomalous properties have not
been elaborated.  Nevertheless, there is some confidence that
it will exhibit topological instability and topological
superconductivity along the scenario outlined in these lectures.

\section{Conclusion}\label{VI}

In summary, we have seen that if the ground state of a quantum system
dynamically acquires a non-zero Chern's number (i.e.\ a topological
order), then it is superconductive.  In units where $e=c=m=1$, the
London depth is literally the density of the topological charge.  This
connection is rigid and as is the case with the whole theory, it is
determined by the geometry of the phase space.  It therefore seems
likely that superfluidity can be derived on a more phenomenological
(model independent) basis.

It is also clear that topological superconductivity is drastically
different from the BCS mechanism.  How does this difference manifest
itself in observables?  For example, what are the tunneling properties
of the topological superconductor, and the symmetry of the order
parameter?  This is perhaps the most interesting avenue of developing
this theory.

\bigskip
\noindent
{\bf Acknowledgement:}
This work is supported in part by NSF Grant STC-8809854 and
by the Japan Society for the Promotion of Science. I acknowledge the help
provided by
A. Abanov in preparation these lectures and also numerous discussion whis
him on different aspects
of topological mechanism of superconductivity.

%\pagebreak
%\references

\end{document}